\documentclass[twocolumn,showpacs,preprintnumbers,amsmath,amssymb,prb]{revtex4-1}
\usepackage{graphicx}
\usepackage{color}
\usepackage{amsmath,amsthm,amssymb}
\usepackage{subfigure}
\begin{document}
 
\title{Self-consistent scheme for optical response of large hybrid networks of semiconductor
quantum dots and plasmonic metal nanoparticles}

\author{L. Hayati\textsuperscript{1,}}\thanks{These two authors contributed equally}
\author{C. Lane\textsuperscript{2,}}\thanks{These two authors contributed equally}
\author{B. Barbiellini\textsuperscript{2}}
\author{A. Bansil\textsuperscript{2}}
\author{H. Mosallaei\textsuperscript{1}}
\affiliation{
\textsuperscript{1}CEM and Photonics Lab, Electrical and Computer Engineering Department, Northeastern University, 
Boston 02115, USA, \\
\textsuperscript{2}Physics Department, Northeastern University, 
Boston MA 02115, USA \\}
\date{version of \today} 
\begin{abstract}
We discuss a self-consistent scheme for treating the optical response of large, hybrid networks of semiconducting quantum dots (SQDs) and plasmonic metallic nanoparticles (MNPs). Our method is efficient and scalable and becomes exact in the limiting case of weakly interacting SQDs. The self-consistent equations obtained for the steady state are analogous to the von Neumann equations of motion for the density matrix of a SQD placed in an effective electric field computed within the discrete dipole approximation. Illustrative applications of the theory to square and honeycomb SQD, MNP and hybrid SDQ/MNP lattices as well as SQD-MNP dimers are presented. Our results demonstrate that hybrid SQD-MNP lattices can provide flexible platforms for light manipulation with tunable resonant characteristics. 
\end{abstract}
\maketitle 

\section{Introduction}

Collective surface charge oscillations (plasmons) on a metal nanoparticle (MNP) can strongly localize light to subwavelength regions and greatly enhance the field in these regions\cite{atwater2007a,abajo2007a,engheta2007a,ciraci2012a,manjavacas2014a,dutta2015a}. Gold nanoparticles, for example, are well known to exhibit plasmonic resonances in the visible\cite{novotny2012a,brongersma2007a}. Hybrid systems of MNPs and semiconductor quantum dots (SQDs)\cite{govorov2006a,curto2010unidirectional,mertens2006polarization,fedutik2007exciton,akimov2007generation,pons2007quenching,wei2014a} are attracting special interest because interactions between the excitons of an SQD and the plasmons of an MNP can lead to novel effects and strong modifications of the optical properties of an SQD-MNP network compared to those of the underlying SQD or MNP building blocks; the SQDs play the role of quantum emitters in the network\cite{brus1986a,murray2000a}, whereas the MNPs act to amplify or dampen the electromagnetic field. The matrix elements of the density operator satisfy the well-known optical Bloch equations\cite{haug1990a}. Thus, as shown by Zhang {\it et al.}\cite{zhang2006a}, the plasmon-excitation interaction leads to the formation of a hybrid excitation with shifted frequency (Lamb shift) and decreased lifetime. The modified decay rate can be also derived  from Fermi's golden rule as shown in the book by Novotny and Hecht\cite{novotny2012a}. Efficient transfer of energy through the network can be achieved by designing a hybrid layer composed of plasmonic elements coupled with  
SQDs\cite{renu2014a} or semiconducting interfaces\cite{knight2011a}. The underlying mechanism involves a near-field resonance of electric dipoles, also known as Forster resonance energy transfer \cite{king2012a}, which can be viewed as a quantum version of the classical resonance phenomenon\cite{ansari2013a}. 

Exciton migration in a hybrid SQD-MNP network can be incoherent (diffusive)\cite{govorov2006a} or coherent (wavelike) and could be studied by using positronium atom simulators in a metal-organic framework.\cite{crivelli2014positronium}
In the coherent case, excitations are transferred back and forth between the MNPs and the SQDs. This regime occurs in the vicinity of the exciton-plasmon resonance and produces a shift in the exciton emission frequency. Coupling of the resonance to the broad continuum of plasmonic modes of the MNP in the presence of an applied driving field near the resonance has been investigated within the framework of the quantum density-martix formalism\cite{zhang2006a,artuso2008a,artuso2010a,artuso2013a}, and is shown to yield Fano lines shapes\cite{Fano1961a,luk2010fano,miroshnichenko2010fano}, excitation induced transparency and suppression, and bistability behavior of the network\cite{zhang2006a,artuso2008a}. Reference \onlinecite{artuso2013a} has developed a theory in this connection, but the scheme of Ref. \onlinecite{artuso2013a} can only treat a few 
building blocks since it involves a set of 
complicated non-linear ordinary differential equations (ODEs). 
In order to address this bottleneck in system size, we consider a set of linear von Neumann equations of motion in the steady state
for the density matrix of each SQD placed in an effective field, due to the network, which is obtained within the discrete dipole approximation (DDA).

The resulting equations differ sharply from the standard linear response treatment in that our SQD density matrix operator can be cast in terms of occupation numbers, which can be computed very efficiently by adapting self consistent field (SCF) iterative schemes that have been implemented in many quantum chemistry and solid-state electronic structure software packages involving Hartree-Fock or Kohn-Sham equations\footnote{Rigorous convergence properties of these schemes are still a field of active research [C. Yang, J. C. Meza, B. Lee, and L.-W. Wang, ACM Transactions on Mathematical Software (TOMS) {\bf 36}, 10 (2009)]}. In this way, our method becomes extremely efficient and scalable and 
enables the treatment of very large hybrid networks. 

The present framework will allow a broader exploration of light-matter interactions in metamaterials\cite{engheta2006a,memarzadeh2013a,alu2006a} and hybrid systems\cite{cheng2013a,valleau2014a,salary2015a}, where  one is constrained currently to the treatment of only plasmonic particles. Inclusion of SQDs, offers a much greater degree of freedom in the development of applications such as multi-wavelength energy absorption arrays\cite{ghadarghadr2009a,cheng2014a} and optical nanocircuits\cite{engheta2007a}.

\section{Method}
\subsection{Formalism}
Our scheme is composed of two main parts, namely, the evaluation of the von Neumann equations of motion 
for the density matrix, $\rho$, of each SQD in the steady state, 
and of the effective electric fields calculated within 
the DDA\cite{purcell1973a,draine1988a, draine1993a,draine1994a,draine2000a,yurkin2011a}.  
The density matrix of each SQD is 
first initialized to the one given by 
the external electric field $E_{0}$. 
It is next updated 
by using the local electric field at the SQD. 
In each of these steps, we solve the master equation for the steady state given by:
\begin{align}
\frac{d\mathbf{\rho}}{dt}&=\frac{i}{\hbar}\left[\mathbf{\rho},\mathcal{H}_{E}\right]-\Gamma(\rho).
\label{eqheis}
\end{align}
In Eq. (\ref{eqheis}) $\mathcal{H}_{E}=\hbar\omega_{0}\hat{a}^{\dagger}\hat{a}-\mathbf{\mu} \cdot \mathbf{E}\hat{a}-\mathbf{\mu} \cdot \mathbf{E}^{*}\hat{a}^{\dagger}$ is the SQD Hamiltonian, 
where $\hat{a}$ and $\hat{a}^{\dagger}$ are the exciton annihilation and creation operators, $\omega_{0}$ is the energy gap in the SQD, 
$\mathbf{\mu}$ denotes the dipole matrix element, and $\mathbf{E}$ is the electric field. Moreover, $\Gamma$ is the relaxation matrix where the matrix elements are $\Gamma_{11}=(\rho_{11}-1)/\tau_{0}$, $\Gamma_{22}=\rho_{22}/\tau_{0}$, and $\Gamma_{12}=\Gamma_{21}^{*}=\rho_{12}\gamma_{21}$. In this work we have taken the values of $\gamma_{21}$ and $\tau_{0}$ from  Ref. \onlinecite{zhang2006a,artuso2008a} in order to benchmark our solution. For including the Purcell effect, decay rates can be renormalized to take into account effects of the environment as discussed in Ref. \onlinecite{novotny2012a,esteban2014strong}.
Here, entanglement effects have been neglected since these are observed to be small in the steady state \cite{hou2014a,dzsotjan2011a}, although these effects can be significant in the transient regime.

In order to find the induced polarizations 
on various 
elements of the hybrid network within the DDA, 
we assign polarization $\mathbf{P}_{i}$ and polarizability $\alpha_{i}=\varepsilon_{0}\chi_{i}$\citep{ahmadi2010a,alu2007a} to the $i^{th}$ element (plasmonic or semiconducting) of the network\footnote {Our use of the DDA is justified when particles are separated by sufficiently large distances so that higher-order modes can be neglected\cite{park2004a,romero2006a}. For a sphere, Ref. \onlinecite{gerardy1982a} considers a scheme going beyond the DDA.}. Then $\mathbf{P}_{i}=\alpha_{i}\mathbf{E}_{loc}^{i}$, 
where $\mathbf{E}_{loc}^{i}$ is the total (local) electric field on the $i^{th}$ site 
produced by all other sites and the external electric field. This expression can be expressed in 
a system of linear equations\cite{rashidi2010a} given by
\begin{align}\label{eq:polmatrix}
P_{x}^{i}=\alpha_{i}\left[\sum_{i\neq j}\left(G_{xx}^{ij}P^{j}_{x}+G_{xy}^{ij}P^{j}_{y}\right) +E^{0}_{x} \right],\\
P_{y}^{i}=\alpha_{i}\left[\sum_{i\neq j}\left(G_{yx}^{ij}P^{j}_{x}+G_{yy}^{ij}P^{j}_{y}\right) +E^{0}_{y} \right],\nonumber
\end{align}
where $P_{x}^{i}$ and $P_{y}^{i}$ are the $x$- and $y$- components of the polarization at the $i^{th}$ site, and 
$E^{0}_{x}$ and $E^{0}_{y}$ are the $x$- and $y$- components of the external electric field. 
$G_{ws}^{ij}$, with $ws\in \{xx,xy,yx,yy\}$, is a matrix element of the dyadic Green's function 
$\overset{\leftrightarrow}{\mathbf{G}}(\mathbf{r},\mathbf{r}^{\prime})$, where $\mathbf{r}$ is the location of the $i^{th}$ observation site and 
$\mathbf{r}^{\prime}$ is the location of the $j^{th}$ source site. The resulting closed form of 
$\overset{\leftrightarrow}{\mathbf{G}}(\mathbf{r},\mathbf{r}^{\prime})$ given in Ref.\onlinecite{chen1983theory} is
\begin{align}
\overset{\leftrightarrow}{\mathbf{G}}(\mathbf{r},\mathbf{r}^{\prime})&=\frac{1}{4\pi\varepsilon_{0}}\frac{e^{-ik_{o} R}}{R^{3}}\left\{  \left[(k_{0}R)^{2}-ik_{0}R-1\right]\overset{\leftrightarrow}{\mathbf{I}}-\right.\nonumber\\
&\left.-[(k_{0}R)^{2}-3i k_{0}R-3]\mathbf{R}\mathbf{R}\right\},
\end{align}
where $\overset{\leftrightarrow}{\mathbf{I}}$ is the identity dyad, 
$\mathbf{R}=\mathbf{r}-\mathbf{r}^{\prime}$, and $k_{0}$ is the free space wave vector.
    		\begin{center}
    		\begin{figure}[h]
     		\includegraphics[clip=true,trim = 34pt 1pt 1pt 1pt,scale=.75]{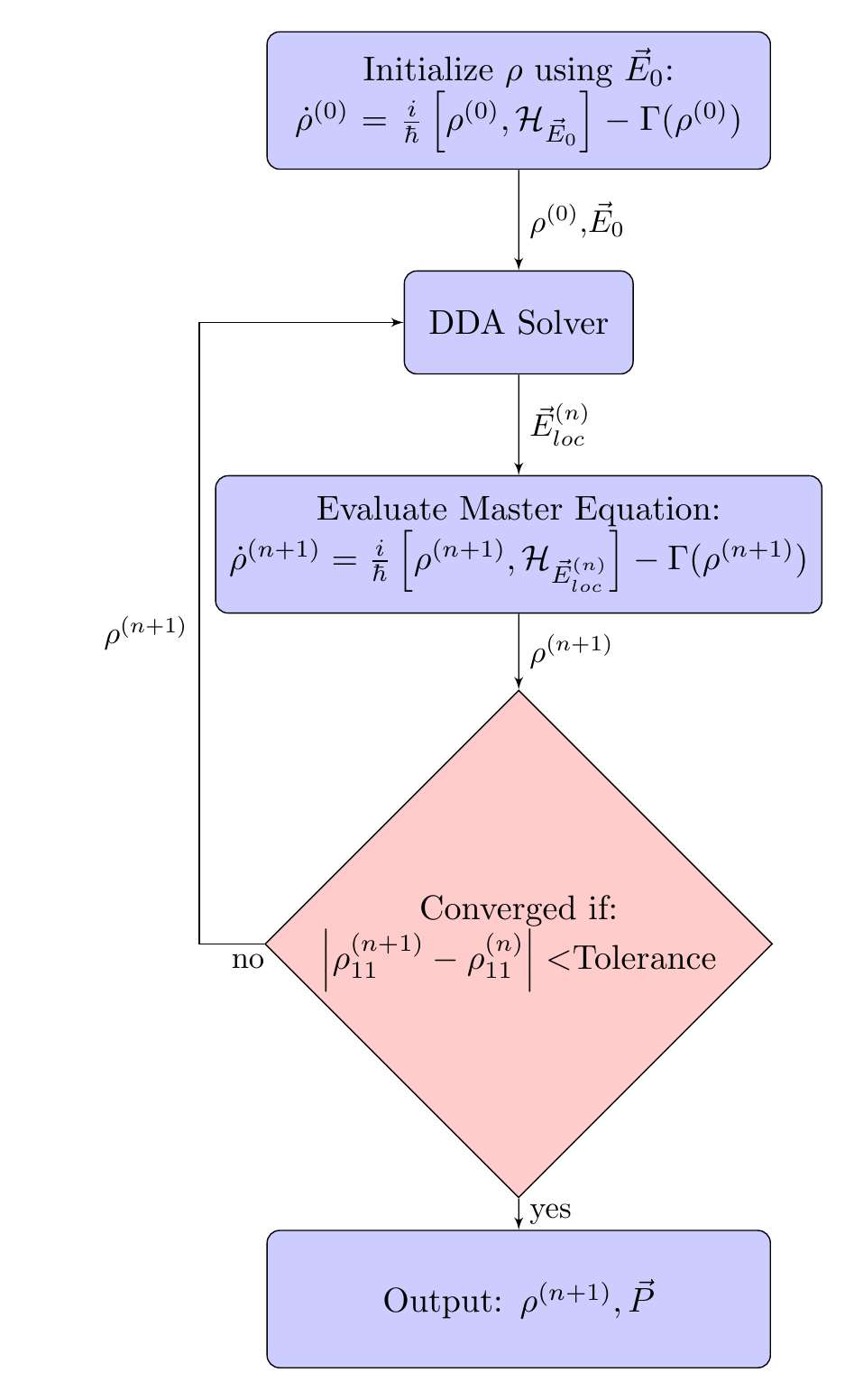} 
			\caption{Schematic illustration of our self-consistency loop for treating the network of SQD and MNP elements. Here $\rho^{n}$ denotes the density matrix of a SQD in the $n^{th}$ iteration.}  
			\label{fig:flowchart}   		
     		\end{figure}
    		\end{center}
    		
Since our network contains two distinct types of elements (MNP and SQD), 
we must consider two different forms of linear susceptibility. 
The classical MNP susceptibility is given by
\begin{align}
\chi_{MNP}&=4\pi\varepsilon_{0}a^{3}\gamma,
\end{align}
where $a$ is the radius and $ \gamma=\frac{\varepsilon_{m}-\varepsilon_{0}}{\varepsilon_{m}+2\varepsilon_{0}}$ 
is the effective dielectric constant of the MNP. 
For the SQD, we use 
\begin{align}\label{eq:qdsusep}
\chi_{SQD}&=\dfrac{1}{3\hbar\epsilon_{eff}}\dfrac{2\omega_{0}\rho_{11}\mu^2}{(\omega_{0}-\omega-\gamma_{12})(\omega_{0}+\omega+\gamma_{12})}
\end{align}
where $\epsilon_{eff}$ is given in Ref.~\onlinecite{zhang2006a}, $\rho_{11}$ is a matrix element of the density matrix, and $1/\gamma_{12}$ is the lifetime of the excited state\citep{boyd2003a} given by the relaxation matrix in Eq. (\ref{eqheis}).
\subsection{Practical Implementation of the Algorithm}
The combined evolution of the density matrix and the induced local polarizations can now be obtained through the preceding set of equations, starting with the initial density matrix $\rho^{(0)}$ and the resulting susceptibility $\chi_{SQD}$ [see Fig. \ref{fig:flowchart}]. The linear system in Eq. (\ref{eq:polmatrix}) is solved self-consistently to 
yield the polarizations on various elements of the network using the local field on each SQD to extract an updated density matrix
$\rho^{(1)}$.  The main computational cost as a function of the size of the system is driven by the matrix inversion of the linear system in Eq. (\ref{eq:polmatrix}) the complexity of which, depending on the algorithm, can range from $O(n^{2}\log{}n)$ to $O(n^{3}$) as shown in Ref.\onlinecite{Cormen:2001:IA:580470}. Here $n=d(N_{MNP}+N_{SQD})$, with $d$ being the number of spatial dimensions and $N_{MNP}$ $(N_{SQD})$ being the number of MNPs (SQDs) in the system. Self-consistency is reached when $\left| \mathbf{\rho}^{\left(n+1\right)}_{11} - \mathbf{\rho}^{\left(n\right)}_{11} \right|$ is smaller
than a given value: here we used a tolerance of $10^{-5}$.
In the present calculations, we found that convergence of the density matrix is typically achieved within about ten iterations, with the number of iterations depending on the external field strength, dipole strength, the distance between the particles and the proximity of the system to the resonance frequency $\omega_{0}$. However, when $\mu$ is large, and the distance between particles is small, we found an increase in the number of iterations to around 30.

    		\begin{center}
    		\begin{figure}[h]
     		\includegraphics[scale=.39]{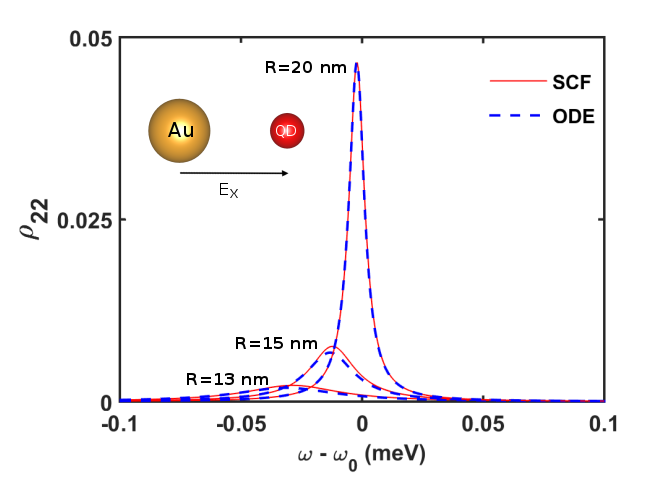}
     		\caption{(color online) Population of the excited state of a dimer system for different inter-particle distances $R$. The ODE and SCF results are compared.}
     		\label{fig:dimer}   		
     		\end{figure}
    		\end{center}
    		
    		\begin{center}
    		\begin{figure}[h]
     		\subfigure[]{\includegraphics[scale=.18]{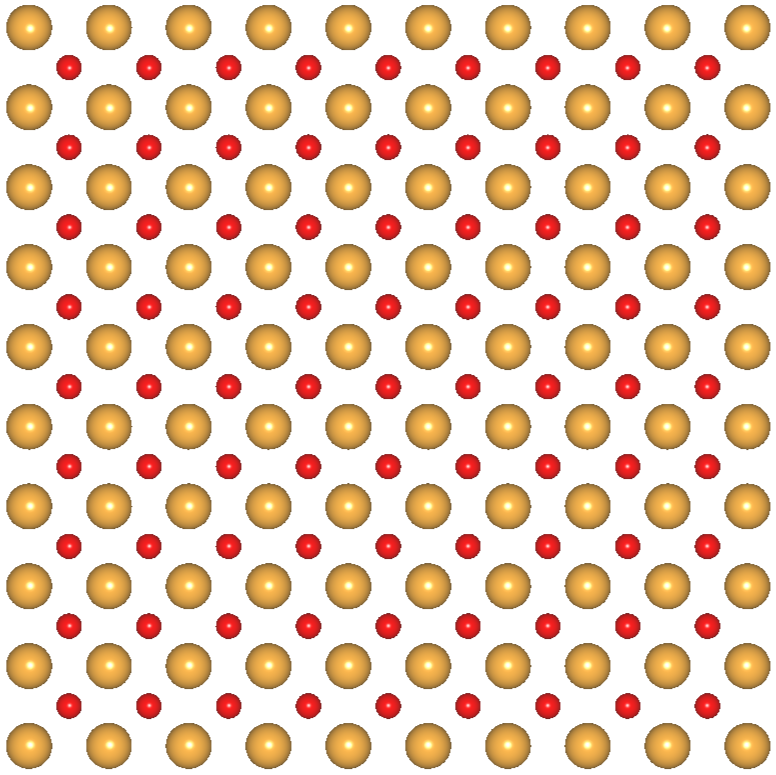}}
     		\hspace{+4pt}
     		\subfigure[]{\includegraphics[scale=.27]{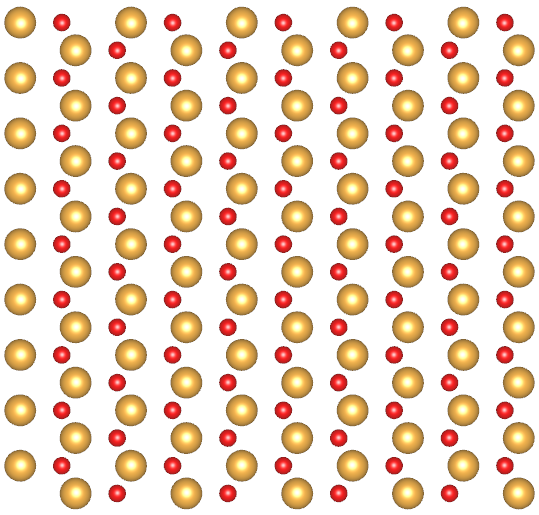}}
     		\caption{(color online) (a) A $10\times 10$ square MNP lattice with a basis of SQDs; (b) a $10\times 10$ MNP/SQD honeycomb lattice. The gold (red) spheres represent the MNP (SQD), and the +x (+y)-axis goes rightward (upward).} 
			\label{fig:structure}   		
     		\end{figure}
    		\end{center}

    		\begin{center}
    		\begin{figure}[h]
     		\includegraphics[scale=.20]{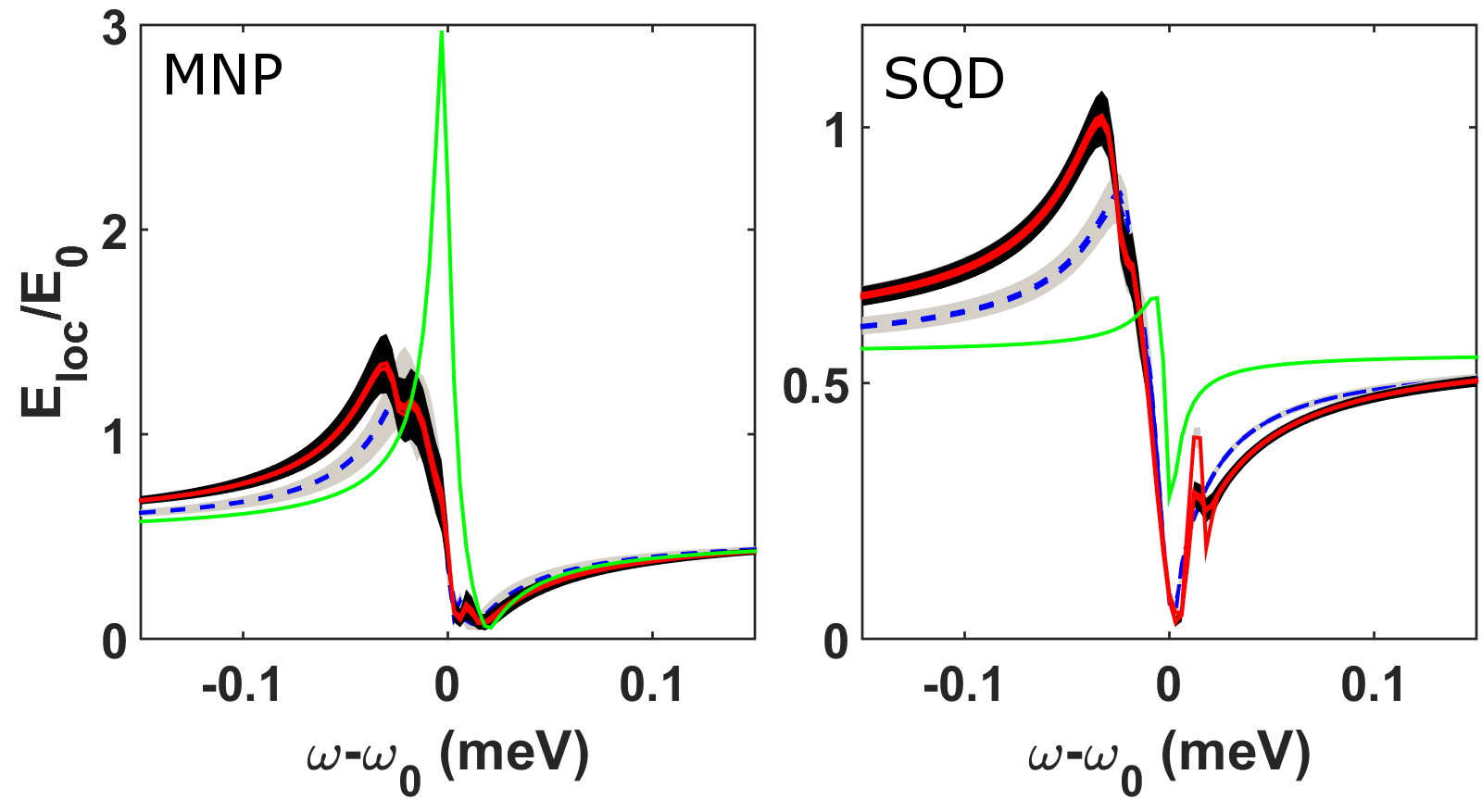}
     		\caption{(color online) Resonant behavior of the local electric field $E_{loc}$ on MNP (left) and SQD (right) elements of various lattices as a function of the frequency $\omega$: square lattice (blue lines); honeycomb lattice (red lines); and dimer case (green lines). The effect of disorder in the lattice is shown by shading of different colors around various lines. The external field is oriented $45^{\circ}$ with respect to the $x$-axis.} 
			\label{fig:field}   		
     		\end{figure}
    		\end{center} 

The standard route followed in quantum plasmonics involves solving simultaneously the
rate equations of the quantum emitters along with the field equations
obtained via finite-difference time-domain schemes. 
The present SCF approach gives the same results in the steady state 
as the standard approach\cite{zhang2006a}. However, our scheme avoids the key numerical bottlenecks of the standard approach by invoking the SCF methodology. It should be noted that a disadvantage of our method is that it cannot treat the transient regime before the system reaches the steady state. If effects of the fractional occupation number $\rho_{22}$ of the excited state are neglected, we don't need to iterate and our scheme becomes equivalent to that introduced by Panahpour {\em et al.}\cite{panahpour2012a}. For a system only composed of SQDs, our SCF scheme reduces to the generalized Maxwell-Bloch equations introduced by Bowden and Dowling\cite{bowden1993near}.

\section{Results}

We first consider the illustrative case of a hybrid dimer composed of a spherical MNP of radius $a=7.5$ nm and an SQD in the presence of a polarized external field  $E_0 \cos(\omega \text{{\it t}})$, at light intensity of $I_0 = 1~\text{W/cm}^2$. Plasmonic properties of the MNP are introduced in our calculations by using the dielectric function of Ref.  \onlinecite{johnson1972a}. The energy gap $\omega_{0}$ of the SQD can be tuned to resonate with the MNP, for example, by modifying the size of the SQD\cite{brus1986a}. The dipole moment of the SQD is given by $\mu=e~r_0$ where we take $r_0 = 0.65$ nm, and the relaxation times to be $\tau_0=0.8$ ns for fluorescence and $1/\gamma_{12}=0.3$ ns for the dipole transition. As we noted earlier the values of $\gamma_{21}$ and $\tau_{0}$ are taken from Ref.\onlinecite{zhang2006a,artuso2008a}. The center-to-center distance between the two nano-particles, $R$, ranges typically between $13$ and $80$ nm. Depending on the angle between the polarization vector and dimer axis, the two dipoles will interfere either constructively or destructively. In particular, the induced field between the spheres will be enhanced in the longitudinal polarization configuration at frequencies well below the resonance. 
Figure 2 shows the population of the excited state, $\rho_{22}$, for the SQD in the dimer system for different inter-particle distances $R$ when the field is in the longitudinal polarization configuration. The earlier ODE results of Refs. \onlinecite{zhang2006a,artuso2008a} are seen to be almost identical to the present SCF results for $R=20$ nm, although one can notice small differences at shorter inter-particle distances. The reason is that our self-consistent computation fully captures the feedback of dipole interactions in the system. In fact, in the small $R$ limit, we find that the MNP dominates the response and the SQD becomes irrelevant, while for large $R$, the behavior of the MNP and SQD contributions is opposite.  Our method thus correctly captures the standard ODE cases of dimer as well as SQD/MNP/SQD \cite{artuso2013a} and MNP/SQD/MNP trimers as shown in detail in the Supplementary Information (SI). Our analysis indicates that for $15\leq R\leq 20$ nm, the hybrid artificial systems (dimer or trimer) behave significantly differently from their constituent elements, and offer unique optical properties at the nanometer scale at their resonant energies. 

In particular, when $\mu$ is large, our method is 
able to capture plexitonic effects such as electromagnetically induced transparency and modified Fano shapes; it also reproduces cases studied with the standard ODE approach by Artuso and collaborators\cite{artuso2008a,zhang2008a}. Interestingly, Artuso {\it et al.} found two distinct solutions to the rate equations\cite{artuso2010a,artuso2008a,artuso2012a} due to non-linearity in the dimer case for a specific set of parameter values ($R=13\text{nm}, a=7\text{nm}, \mu=3.5 e~\text{nm}$). One of these stable solutions is a smooth and continuous function of $\omega$, while the second solution displays a similarly broad asymmetrical shape away from the resonance with a discontinuous jump. Our method, on the other hand, only yields the first solution. In the strong-coupling regime discussed in Ref. \onlinecite{esteban2014strong}, the atom-field coupling $\kappa$ (see Ref.\onlinecite{novotny2012a} for definition) is much larger than the spontaneous decay rate. 
Such a regime can be accessed by measuring vacuum Rabi oscillations\cite{vasa2013real}.

			 \begin{center}
    		\begin{figure}[h]
     		\includegraphics[scale=.16]{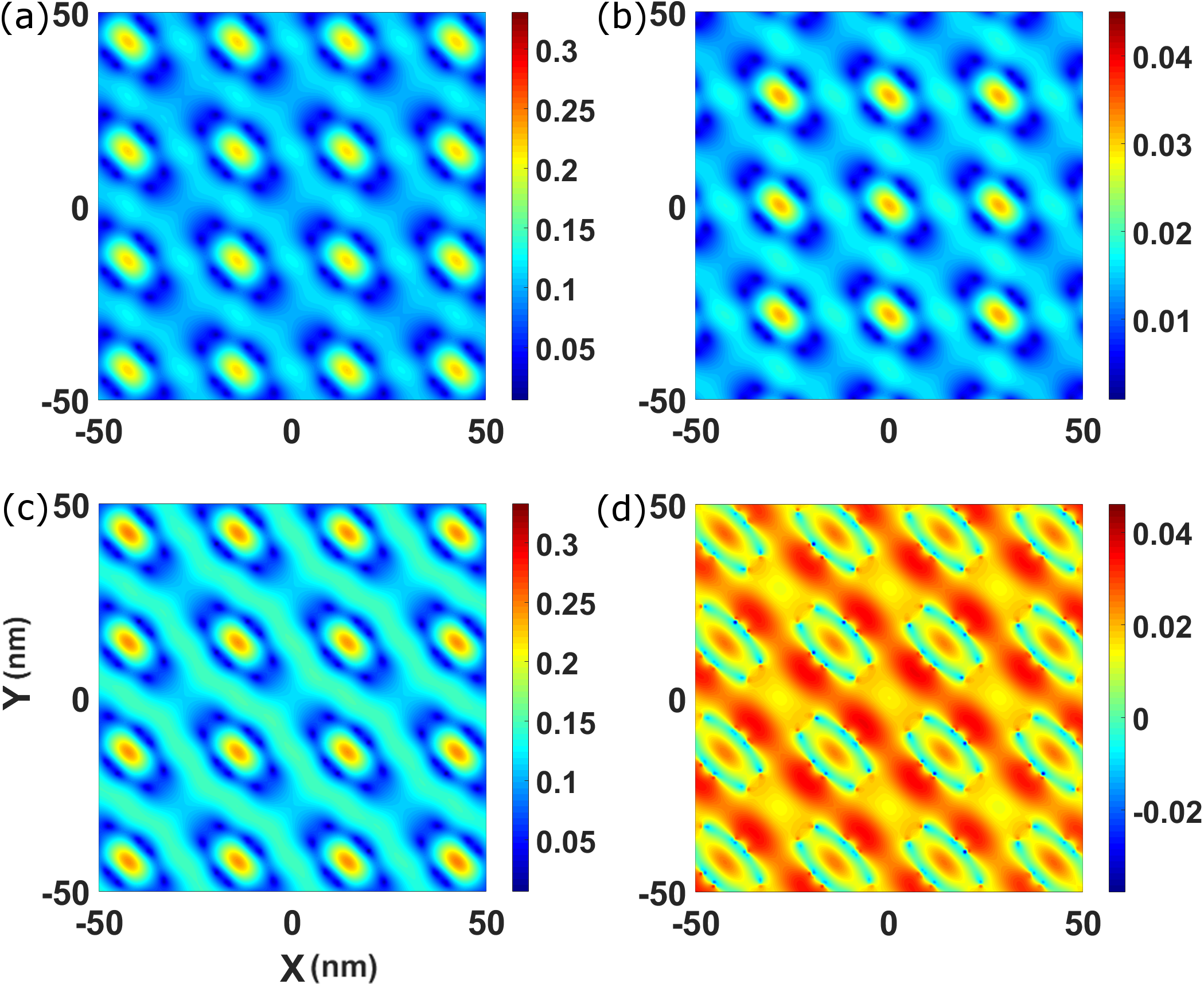}
     		\caption{(color online) Intensity of the induced electric field (excluding the external field $E_{0}$) in a plane $12$ nm above the square 10x10 planer network for: $(a)$ a pure MNP network; $(b)$ a pure SQD network; $(c)$ the hybrid MNP/SQD network, and, $(d)$ the difference between the hybrid system in $(c)$ and the pure MNP system in $(a)$. The external field is oriented at $45^{\circ}$ with respect to the $x$-axis with $\omega$ at resonance. The field intensities are given in units of the external field intensity.} 
			\label{fig:focal}   		
     		\end{figure}
    		\end{center} 
 
We turn now to discuss the electromagnetic response of hybrid SQD/MNP lattices by taking advantage of the high computational efficiency of our SCF algorithm. Properties of two specific lattices are considered: a $10\times 10$ square MNP lattice with a basis of SQDs at $(0.5,0.5)$, and a $10\times 10$ MNP/SQD honeycomb lattice, see Fig.\ref{fig:structure}. 
Such large systems are intractable within the standard ODE approach\cite{artuso2012a}. 
In investigating the SQD/MNP networks, we chose $R=20$ nm as the distance between the SQD and MNP 
elements for ease of comparison with the corresponding dimer results. Figure~\ref{fig:field} illustrates 
the resonant behavior of the local electric field $E_{loc}$
as a function of the frequency $\omega$ of 
the external electric field, which is oriented $45^{\circ}$ with respect to the $x$ axis. 
We see that on the SQD site of the square lattice there is a strong suppression of the local electric field at the resonance frequency [blue curve in Fig. \ref{fig:field}(b)] and just before the resonance $E_{loc}$ becomes larger than $E_{0}$. In the honeycomb lattice also the ratio $E_{loc}/E_{0}$ rises just before the resonance but it does not become larger than unity. By comparing various curves in Fig. \ref{fig:field}, it is clear that there are substantial differences between the behavior of the SQD and MNP lattices, and that the response of the lattices differs sharply from that of the dimer, especially at and near the resonance. Results of Fig. \ref{fig:field} demonstrate that the $E_{loc}/E_{0}$ line shape can be controlled through the choice of the lattice on which elements of the network are arranged, providing flexibility in tuning the plasmonic characteristics of the network. 
We have also taken advantage of the scalability 
of our algorithm to find that, near the resonance frequency,
the density operator in the infinite lattice limit needs
systems as large as $80\times 80$ 
to converge as illustrated in the SI\footnote{See Fig.S2 in the SI.}.  
Finally, we have simulated effects of disorder 
by randomly varying the positions of the SQDs and MNPs 
in the lattice by up to 5\% of the inter-particle distance away from the perfect lattice positions. The resulting uncertainty in the response is shown by the shading around various curves in Fig. \ref{fig:field}. It is seen that the response in all cases considered in Fig. \ref{fig:field} is quite robust against such disorder effects.

Figure~\ref{fig:focal} gives further insight into our results by 
showing that the  hybrid network can be 
used to shape the electric field in the near-field 
region by producing a beam with a modulated pattern.  
Here, we consider the $10\times 10$ MNP/SQD square network 
discussed above using the same external field orientation.
Figure~\ref{fig:focal}(a) shows the electric field in a plane $12$ 
nm above the planar network for the SQD subnetwork, which is compared 
with the corresponding results of Fig. \ref{fig:focal}(b) for 
the MNP subnetwork\cite{rashidi2010a}. The focal properties of 
the full hybrid MNP/SQD system (panel $(c)$)
are seen to change significantly as demonstrated by the 
difference, panel $(d)$, with respect to the linear superposition 
of the two pure systems (i.e. MNP and SQD)
\footnote{For the sake of clarity, 
the intensity pattern in the restricted region of size $100$ 
x $100$ nm only is shown.}.
SQD/MNP arrays could thus provide a flexible basis for designing platforms 
for nano-antenna light manipulation\cite{akselrod2014a}.
We have noted above that Fig. \ref{fig:field}
is little affected by randomness. However, the effects of disorder are mainly manifested in
the propagation properties. Therefore, 
quantities shown in Fig.~\ref{fig:focal}, 
which are relevant to
propagation and Green's tensors, 
are much more sensitive to disorder effects as
shown in the SI\footnote{See Fig.S4 in the SI.}. 
Interestingly, disorder in the lattice can also lead to Anderson localization effects as shown by John \cite{john1987strong}, although our main reason to introduce small disorder is to assess the stability of our numerical solutions.

\section{Conclusion}
We have developed an efficient SCF method based on the DDA 
for obtaining the optical response of large networks of plasmonic MNPs and SQDs. 
Our method is both accurate and scalable, 
and it can be generalized to treat complex nano-resonators 
with arbitrary shapes\cite{yang2015a}. 
The present scheme solves the problem of computational bottlenecks for the numerical treatment of large hybrid networks of MNPs and SQDs, and advances the field of opto-electronics based on plasmonics.
For example, one could address in this way the development of optimal architectures for absorbing layers in novel quantum dot sensitized solar cells \cite{renu2014a}. 
By combining MNPs 
with quantum emitters such as the SQDs, it will become possible to model wireless networks at the nanoscale, and analyze the efficiency of energy transport through such networks.

\section{Acknowledgments}
We are grateful to G.W. Bryant for useful discussions on the Purcell effect and for sending us Ref.\onlinecite{esteban2014strong}.
This work was supported by the US Army Research Office grant number W911NF-15-1-0138, and benefited from the allocation of computer time at Northeastern University's Advanced Scientific Computation Center.
\bibliography{mybib}

\end{document}